# Exceptional points in topological edge spectrum of PT symmetric domain walls

Xiang Ni[1,2], Daria Smirnova[1,3,4], Alexander Poddubny[4-6], Daniel Leykam[7], Yidong Chong[7,8], and Alexander B. Khanikaev[1,2*]

[1]*Department of Electrical Engineering, Grove School of Engineering, The City College of the City University of New York, 140th Street and Convent Avenue, New York, NY 10031, USA*
[2]*Graduate Center of the City University of New York, New York, NY 10016, USA*
[3]*Institute of Applied Physics of the Russian Academy of Sciences, Nizhny Novgorod 603950, Russia*
[4]*Nonlinear Physics Centre, Research School of Physics and Engineering, Australian National University, Canberra, ACT 2601, Australia*
[5]*ITMO University, St. Petersburg 197101, Russia*
[6]*Ioffe Institute, St. Petersburg 194021, Russia*
[7]*Division of Physics and Applied Physics, School of Physical and Mathematical Sciences, Nanyang Technological University, Singapore 637371,Singapore*
[8]*Centre for Disruptive Photonic Technologies, Nanyang Technological University, Singapore 637371, Singapore*

*e-mail: akhanikaev@ccny.cuny.edu

**Abstract**: We demonstrate that the non-Hermitian parity-time (PT) symmetric interfaces formed between amplifying and lossy crystals support dissipationless edge states. These PT edge states exhibit gapless spectra in the complex band structure interconnecting complex-valued bulk bands as long as *exceptional points* (EPs) of edge states exist. As a result, regimes exist where the edge states can spectrally overlap with the bulk continuum without hybridization, and leakage into the bulk states is suppressed due to the PT symmetry. Two exemplary PT symmetric systems, based on valley and quantum hall topological phases, are investigated, and the connection with the corresponding Hermitian systems is established. We find that the edge states smoothly transit to the valley edge states found in Hermitian systems if the magnitude of gain/loss vanishes. The topological nature of the PT edge states can be established within the non-Hermitian Haldane model, where the topological invariance is found to be unaffected by gain or loss. Nonreciprocal PT edge states are discovered at the interfaces between PT-Haldane phases, indicating the interplay between the gain/loss and the magnetic flux. The proposed systems are experimentally feasible to realize in photonics. This has been verified by our rigorous full-wave simulations of edge states in PT-symmetric silicon-based photonic graphene.

Subject Areas: Topological Insulator, Photonics, Quantum Physics

## I. INTRODUCTION

Non-Hermitian (NH) Hamiltonians having PT symmetry—the combination of inversion/parity symmetry (P) and time reversal symmetry (TR)—were first systematically examined by Bender and Boettcher [1], who showed that a general class of PT-symmetric Hamiltonians possess real spectra [2]. A region of a system's parameter space where the wave functions are simultaneously eigenstates of both Hamiltonian and PT operators, and hence the spectra are real, is referred to as a "PT-symmetric phase" [3,4]. PT-symmetry can undergo spontaneous symmetry breaking, which causes the spectra to acquire complex-conjugated pairs of eigenvalues. A transition point between a PT-symmetric phase and a symmetry-broken phase is called an exceptional point (EP); here, multiple (usually two) eigenvalues and their associated eigenstates coalesce, and the Hamiltonian becomes defective [5-7]. EPs have been experimentally observed in optics using photonic slabs [8] and waveguides [9], as well as microcavity polaritons [10]. In the past several years, there has been great progress in using photonics to realize and exploit the properties of PT symmetric systems. PT-symmetric optical devices [11] have been shown to exhibit many intriguing characteristics, including beam oscillation in the PT-symmetric optical lattice [12-14], anisotropic transmission resonances [15] in one dimensional (1D) PT-symmetric system, unidirectional invisibility induced by the PT-symmetric refractive index distribution [16] and its experimental demonstration [17,18], single-mode optical [19,20] and acoustic lasers [20,21], nonreciprocal light transmission in silicon photonic circuit [22] and in PT-symmetric whispering-gallery microcavities [23], and optical solitons observed in PT-symmetric coupled fibre loop platforms [24].

One of the most fascinating theoretical questions being explored in the recent PT symmetry literature is what effect non-Hermiticity, and PT symmetric gain/loss in particular, has on topological edge states. The concept of topological edge states of light [25-31] was initially inspired by the studies of quantum Hall and quantum spin Hall phases in Hermitian condensed matter systems [32]. A topological insulator, whether in the electronic or photonic context, possesses a band gap in the bulk, which is spanned by protected surface states [33-35]. However, the established topological classification of band structures, and the attendant bulk-edge correspondence relations giving rise to topological edge states, were formulated for Hermitian Hamiltonians. To the best of our knowledge, to date there remains no general topological classification of NH band structures comparable to the Hermitian topological classifications. New approaches, accounting for the non-Hermiticity in a non-perturbative fashion, are thus highly desirable. Some intriguing connections between PT symmetry and band topology have been discovered. So far, attention has been mostly focused on one-dimensional (1D) models related to the Hermitian Su-Schrieffer-Heeger (SSH) model [36,37]. Such models have been shown to support non-Hermitian variants of the SSH midgap states [38], as well as "anomalous" edge states that are intrinsically non-Hermitian [39]. A few works have also discussed the topological features behind the two-atom chains [40]. The topological stability of edge states in 2D NH systems has been analyzed [41]. Some works have sought to formulate topological invariants for NH models [42,43], including a definition of bulk topological invariants for NH Chern-like topological insulators [43]. The present work attempts to further clarify the effect of gain and loss and PT symmetry on the edge states and topology in general by considering the effect of PT symmetric topological interfaces.

Two-dimensional (2D) honeycomb lattices represent a natural setting for investigating these issues, because both their topological properties in the Hermitian regime and their bulk behaviors in the PT symmetric NH regime are well understood. A honeycomb lattice with unbroken time-reversal (TR) and sublattice symmetries (SS) possesses a band structure with Dirac-like conical dispersion, centered on a pair of "Dirac points" at the ($K$ and $K'$) corners of the Brillouin zone [44]. In addition to Tamm states which occur within the band gaps originating due to Bragg diffraction, hexagonal photonic lattices may exhibit another class of states which appear either in the continuum [44,45] or within the band gaps opened by TR [32] and/or SS reduction without violating Hermiticity [46]. In the Hermitian case, the edge states arise from a combination of lattice symmetries and the topology of each Dirac cone, and it's non-obvious that these features carry over into the NH case. On the other hand, adding gain and loss distorts the Dirac cones, converting them into tachyon-like hyperboloids [47,48]. Hu and Hughes[49] used the symmetry properties of 2D Dirac-type Hamiltonians to argue that nontrivial topology and "strict" PT symmetry are incompatible, in the sense that topologically nontrivial bulk states and their associated topological edge states cannot have completely real spectra after PT symmetric NH terms are added to the Hamiltonian. However, it appears to be possible to evade this restriction under various circumstances [50,51]. In particular, Harari et al. [51] have recently found that purely real unidirectional edge states can appear in a special class of honeycomb lattices – finite temporally modulated (Floquet) lattices with a specific (armchair-only) choice of edges.

Here, instead of periodic modulating lattices in time, we study several simple models of non-Hermitian 2D honeycomb lattices containing two domains with gain and loss, separated by the interfaces formed by zigzag, bearded, or armchair edges. Lossless edge states preserving PT symmetry phase are found located at PT symmetric interfaces irrespective of the cut shape, among which the extensively studied valley edge states are the special example. These PT edge states are more sensitive to the local symmetry of the domain wall than the global symmetry of the structure. If the local parity symmetry at the domain wall is broken, the PT edge states exist only for certain values of the lattice parameters. If local parity symmetry is respected at the domain wall, then PT edge states are robust to the strength of the gain and loss and always exist within the complex-valued bulk bands. Interestingly, instead of relying on the conventional bulk-edge correspondence, the existence of PT edge states can be judged by evaluating the position of their EPs. Thus, guided by this special bulk-edge correspondence in the NH scenario, the presence of EPs of PT edge states ensures the edge bands forming loop and crossing the bulk bandgap both in real and imaginary energy directions by going through these EPs, and the absence of EPs indicate PT edge states either disappear or are gapped out from complex-valued bulk bands. Even more intriguing is that the open gap in the real energy direction is not even required for the presence of the lossless edge states, which makes PT edge states embedded within projected real bulk spectrum without hybridization.

The paper is organized as follows. First, we show that for a valley-Hall-like lattice, there exist real-energy edge states that form loops in the complex energy diagram, bridging the two valleys of the Brillouin zone. In the limiting case of vanishing gain and loss, these "PT edge states" reduce to conventional valley edge states (Sec. II). They are sensitive to the local domain wall symmetry, but robust to the strength of the gain and loss for the case of a locally parity-symmetric wall. Next, the interplay of non-Hermiticity and topology with broken TR symmetry in the context of a non-Hermitian variant of the Haldane model with and without PT symmetric interface is analyzed (Sec. III) and nonreciprocal PT edge states located at the PT symmetric interface are observed. The robustness of PT symmetry is demonstrated by the fact that topological edge states at ends of chain vanish but PT edge states survive beyond the critical value of gain and loss. To test our analytic predictions in experimentally feasible context, an optical analogue of graphene with and without PT symmetric interface is studied both by rigorous full-wave simulations and within the analytical continuous $kp$-type plane wave approximation (Sec. IV).

## II. BRIDGING VALLEYS BY PT EDGE STATES

We first consider a 2D honeycomb valley-Hall lattice model shown in Fig. 1. The lattice consists of two domains, with gain (loss) for site A (B) in the upper domain I and loss (gain) for site B (A) in the lower domain II. The structure has a strip geometry: it is periodic along the $x$ direction, parallel to the interface, and has a finite width $2(N + 1)a_0$ along the $y$ direction with zigzag cut at the ends, where $a_0$ is the lattice constant. Onsite perturbed potentials are also introduced for site A (B) in domain I and site B (A) in the domain II, as shown in Fig. 1(a) (Fig. 1(b)). For the cases of bearded cut and armchair cut in Fig. 1 (c) and (d), the orientation of the strip is the same as that of zigzag cut. Equations of motion are derived from the following tight binding model (TBM) (for details, refer to APPENDIX A):

$$\epsilon\psi_{I,A}(n) = -\psi_{I,B}(n+1) - g_k\psi_{I,B}(n) - \Pi m\psi_{I,A}(n), n = 0,1,2,\ldots,N-1,$$
$$\epsilon\psi_{I,B}(n) = -\psi_{I,A}(n-1) - g_k\psi_{I,A}(n) - \Pi_0 m\psi_{I,B}(n), n = 1,2,\ldots,N;$$

$$\epsilon\psi_{II,A}(n) = -\psi_{II,B}(n-1) - g_k\psi_{II,B}(n) - \Pi_0 m^*\psi_{II,A}(n),\ n = 1,2,\ldots,N,$$
$$\epsilon\psi_{II,B}(n) = -\psi_{II,A}(n+1) - g_k\psi_{II,A}(n) - \Pi m^*\psi_{II,B}(n), n = 0,1,2,\ldots,N-1. \quad (1)$$

Here, $g_k = 2cos(k_x/2)$, $k_x$ is the momentum vector along the $x$ direction, $\psi_{s,j}(n)$ is the component of the wave function within the domain $s = I, II$ at site $(n, j)$, $j = A, B$. $m = m_r + im_i$, $m_r$ is the onsite perturbed potential, and $m_i$ is the magnitude of gain/loss. For the sake of generality, we consider two cases that are both PT-symmetric but differ by the microscopic structure of the interface between the domains. We call these configurations "locally P-symmetric" and "locally P-broken" domain walls, and they are are shown in Fig. 1(a) and Fig. 1(b), respectively. In the first case, $\Pi = 1, \Pi_0 = 0$, the local parity of the sites at the boundary (red rectangle in Fig. 1(a)) is preserved, and the on-site energies adjacent to the wall are real. In the second case, $\Pi = 0, \Pi_0 = 1$, the local parity at the boundary is broken, while the adjacent on-site energies are imaginary. Globally, both domain wall configurations are PT-symmetric.

At the domain wall, the TBM equations are

$$\epsilon \psi_{I,B}(0) = -\psi_{II,A}(0) - g_k \psi_{I,A}(0) - \Pi_0 m \psi_{I,B}(0)$$
$$\epsilon \psi_{II,A}(0) = -\psi_{I,B}(0) - g_k \psi_{II,B}(0) - \Pi_0 m^* \psi_{I,A}(0) \qquad (2)$$

While at the outer boundaries of the strip

$$\epsilon \psi_{I,A}(N) = -\psi_{II,B}(N) - g_k \psi_{I,B}(N) - \Pi m \psi_{I,A}(N)$$
$$\epsilon \psi_{II,B}(N) = -\psi_{II,A}(N) - g_k \psi_{II,A}(N) - \Pi m^* \psi_{II,B}(N). \qquad (3)$$

In Fig. 2, we show the effect of gain and loss on the complex band structure. The complex energies are calculated from Eqs. (1-3) for different values of gain/loss parameter $m_i$ but the same $m_r = 0.3$. Real-valued energies of discrete edge states are found for both locally P-symmetric and P-broken domain walls, and shown in Fig. 2 by thick cyan and red lines, respectively. In the case when $m_r$ is much larger than $m_i$, shown in Fig. 2(a), these lossless edge states look much like the conventional valley edge states. Four edge states are embedded into the bulk spectrum and continuous along $k_x$. If $m_i$ is increased and becomes comparable with $m_r$ ($m_i = m_r = 0.3$ in Fig. 2(b)), the dispersion curves of the edge bands form two heart-shaped loops that are different in size. The edge states for the large loop correspond to the locally P-symmetric domain wall, while those for small loop correspond to the locally P-broken domain wall. If $m_i$ increases to become much larger than $m_r$, the smaller loops shrink and eventually vanish, while the larger loops persist (Fig.2(c)). Lossless edge states for the locally P-symmetric domain wall survive even for very strong gain/loss. Interestingly, we see in Fig. 2(d) that when $m_i$ is large enough to split the loop bands into separate bands, the lossless edge bands persist, embedded within the bulk continuum.

It is interesting to take a closer look at the band structures in the 3D complex space. Figure 3 presents the same cases as in Fig. 2, with the imaginary part of energies being plotted in the 3$^{rd}$ dimension. Because of the PT symmetry of the Hamiltonian, the complex bulk bands have inversion symmetry with respect to the $\epsilon_i = 0$ plane. When $m_i$ is small compared to $m_r$, both the edge states and a few bulk states have real energies. If the magnitude of $m_i$ increases, these bulk states undergo a PT-breaking transition and split into complex conjugated pairs. Only the dispersion curves of the edge states along the domain wall remain real-valued, lying in the $\epsilon_i = 0$ plane [Fig.3 (b-d)]. We note that the looped edge dispersion curves in Fig. 3(b-c) are connected to the bulk modes by parabolic complex-valued dispersion curves. For a zigzag cut, the parabolic connections between edge and bulk states will disappear if the magnitude of $m_i$ is too small or too large, as shown in Fig.3 (a) and (d).

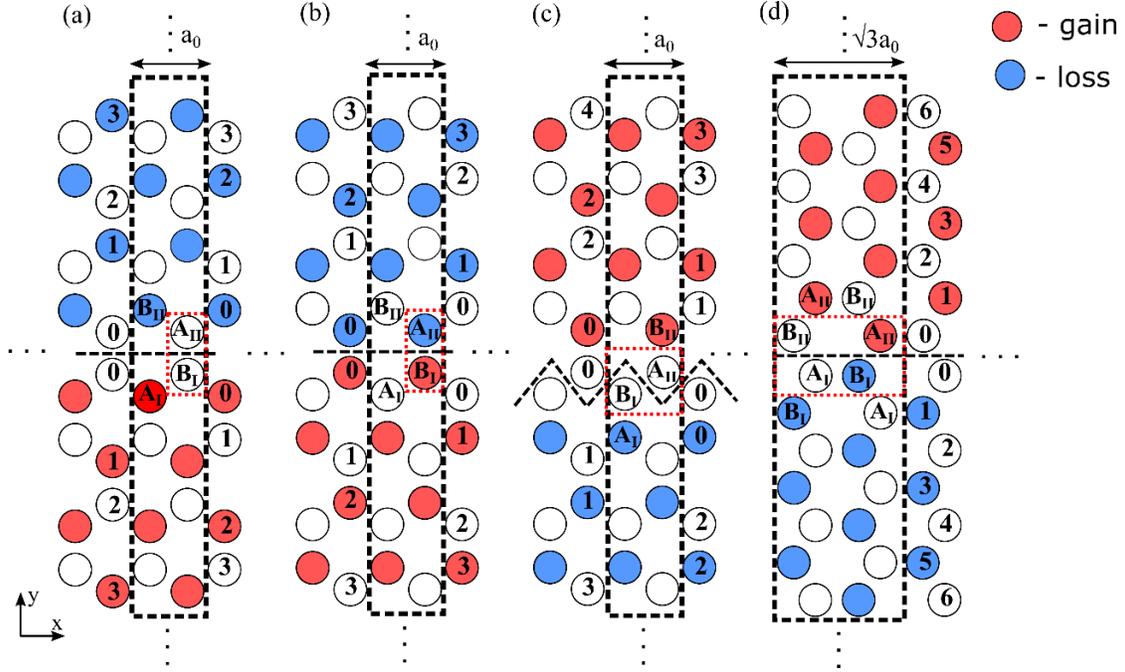

FIG. 1. Different types of PT-symmetric interfaces (valley-Hall domain wall), with different local symmetry of the wall (see the red-dashed rectangle) (a) Zigzag cut with real (locally parity preserved) domain wall. (b) Zigzag cut with imaginary (locally parity broken) domain wall. (c) Bearded cut with real domain wall. (d) Armchair cut with real left part of domain wall and imaginary right part of the domain wall.

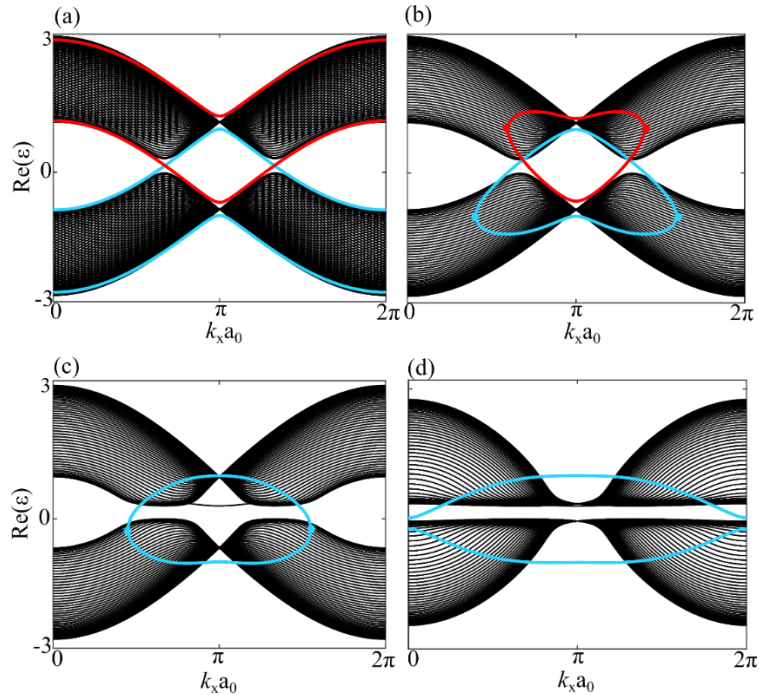

FIG. 2. Energy spectra (black color) calculated from the tight binding model and edge states with found analytically for locally P-symmetric (cyan) and locally P-broken (red) zigzag domain walls. The parameters are (a)

$m_r = 0.3, m_i = 0.05$. (b) $m_r = 0.3; m_i = 0.3$. (c) $m_r = 0.3, m_i = 1.2$. (d) $m_r = 0.3, m_i = 3$. Number of cells for each domain is $N = 50$.

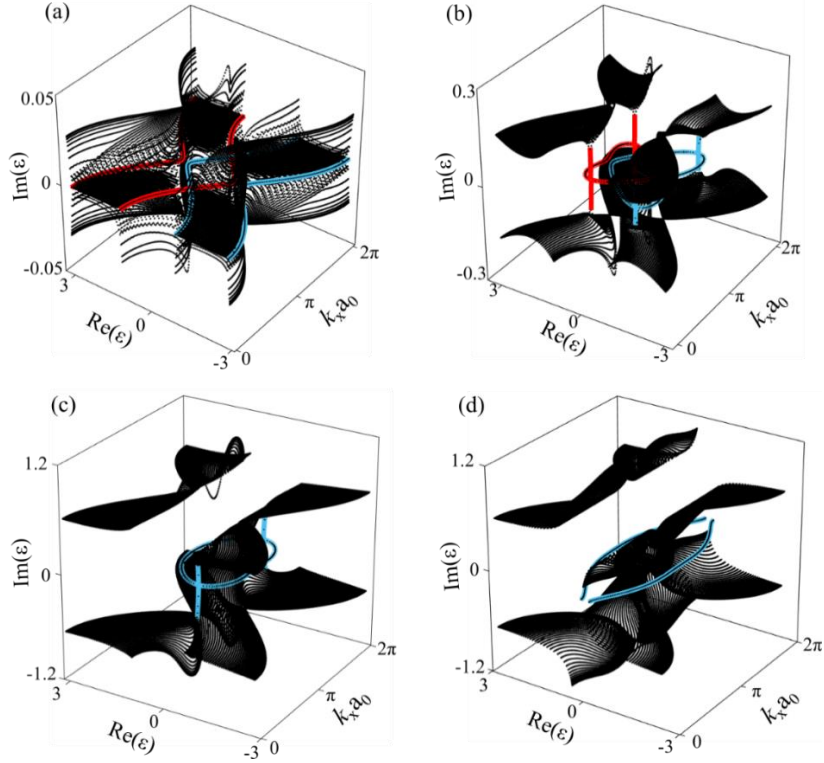

FIG. 3. Complex band structure in 3D for the same cases as in Fig.2.

**PT edge states**

Next, we derive analytic descriptions for the real-energy edge states along the PT symmetric interfaces, which we will henceforth refer to as 'PT edge states'. We start from the equations of motion (1,2). The stripe is considered finite, which formally implies the following boundary conditions at the external boundaries of the stripe

$$\psi_{I/II,B/A}(N+1) = 0. \tag{4}$$

Clearly, for $N \to \infty$ any edge states localized at the ends of the stripe barely feel the effects of gain/loss in the other domain; therefore, they possess complex energies with imaginary parts equal to the magnitude of the gain/loss in their respective domains. Here, we focus on the edge states confined to the central domain wall, whose properties are inherently related to the PT symmetric configuration of the structure.

We observe that the Hamiltonian constructed from Eqs. (1-2, 4) remains invariant under the action of PT symmetry operator defined upon the wave functions as

$$PT\psi(y) = \psi^*(-y). \tag{5}$$

Consequently, if the eigenstates of the Hamiltonian are simultaneously the eigenstates of the PT-symmetry operator,

$$PT\psi(y) = e^{-i\varphi}\psi(y). \tag{6}$$

where $e^{-i\varphi}$ are the eigenvalues of the PT operator, then the eigenvalues of Hamiltonian corresponding to such eigenstates are real [1]; and these states possess the specific symmetry [4]. If the PT symmetric phase is spontaneously broken by tuning the Hamiltonian parameters, the energy eigenvalues are divided into complex conjugate pairs after their states coalesce at the EPs [7] (but still merge with bulk continuum). Both extended and localized states may or may not have PT symmetry phase, and PT symmetry phase is broken for all extended states if the gain/loss is tuned to be large.

Based on Eqs. (5-6), the wavefunction components in the two domains should be related as

$$\psi_{I,A/B}(n) = e^{i\varphi}\psi_{II,B/A}^{*}(n). \tag{7}$$

Thereby, we recover the relation

$$e^{-i\beta} = \frac{\psi_{e,I,A}(n)}{\psi_{e,I,B}(n+1)} = \frac{\psi_{e,II,B}^{*}(n)}{\psi_{e,II,A}^{*}(n+1)} = \frac{\psi_{e,II,A}(0)}{\psi_{e,I,B}(0)} = \frac{\psi_{e,I,B}^{*}(0)}{\psi_{e,II,A}^{*}(0)}, \tag{8}$$

where we have introduced another phase factor $\beta = \varphi - 2\arg(\psi_{e,II,A}(0))$. Eqs. (8) equivalently yield

$$\psi_{e,I,A}(n) = e^{-i\beta}\psi_{e,I,B}(n+1)$$

$$\psi_{e,II,A}(n+1) = e^{-i\beta}\psi_{e,II,B}(n), n = 0,1,\ldots,N-1. \tag{9}$$

The edge states satisfying Eqs. (7) belong to the PT symmetric phase, and the corresponding energy spectra are real. PT edge states are supposed to be localized at the domain wall; moreover, they are concentrated at sites $(0, B)$ in domain I and sites $(0, A)$ in domain II.

Thus, the solutions for the edge states assume the Bloch form

$$\psi_{e,I,A}(n) = a_I e^{ik_{y,I}(n+1)}, \psi_{e,II,A}(n) = a_{II} e^{ik_{y,II}n}$$

$$\psi_{e,I,B}(n) = b_I e^{ik_{y,I}n}, \psi_{e,II,B}(n) = b_{II} e^{ik_{y,II}(n+1)}, n = 0,1,\ldots,N-1. \tag{10}$$

where, due to the PT symmetry condition Eq. (7), the wave vectors and Bloch function amplitudes are related as $k_{y,I} = -k_{y,II}^{*} = p + i\kappa$, $a_I = e^{i\varphi}b_{II}^{*}$, $b_I = e^{i\varphi}a_{II}^{*}$. The parameter $\kappa^{-1}$ characterizes the decay length away from the interface. Remarkably, utilizing the Bloch ansatz Eq. (10) in Eqs. (8) with the boundary Eqs. (2,3), we get the continuity condition for the Bloch vector components $a_I = a_{II}$, $b_I = b_{II}$, being of the same absolute value

$$\frac{a_I}{b_I^{*}} = \frac{a_{II}}{b_{II}^{*}} = \frac{a_I}{b_{II}^{*}} = \frac{b_I}{a_{II}^{*}} = e^{i\varphi}.$$

Substituting Eq. (10) into Eq. (1), we then obtain

$$\begin{bmatrix} (\epsilon_e + m + e^{i\beta})e^{-ik_{y,I}} & g_k \\ g_k & (\epsilon_e + e^{-i\beta})e^{ik_{y,I}} \end{bmatrix} u_e = 0, \tag{11}$$

where $u_e = [a_I, b_I]^T$. Solving the secular equation Eq. (11) and separating the real and imaginary parts, we get two equations, which define the dispersion of PT edge states

$$m_i(\epsilon_e + \cos(\beta)) + m_r \sin(\beta) = 0,$$

$$(\epsilon_e + m_r)\epsilon_e + 2\cos(\beta)\epsilon_e + m_r \cos(\beta) - m_i \sin(\beta) - g_k^2 + 1 = 0. \tag{12}$$

Alternatively, denoting the ratio of real and imaginary parts of the mass term $r = m_r/m_i$, we rewrite Eq. (12) as

$$(r^2 - 1)\epsilon_e^2 + (1 + r^2)m_r\epsilon_e - (1 + r^2)(g_k^2 - 1) = \pm(2r\epsilon_e + m_r(r + r^{-1}))\sqrt{-\epsilon_e^2 + r^2 + 1} \,. \quad (13)$$

The analytically derived dispersion of the edge modes perfectly agrees with the numerical tight-binding calculations.

Remarkably, the parity symmetry with respect to the interface is restored if no gain/loss is present at the lattice sites, i.e. $m_i = 0$. Consequently, the phase difference may take two values $\beta = 0, \pi$, which clearly corresponds to the symmetric and anti-symmetric wave functions of the Hermitian valley edge states, respectively [52]. Therefore, the *valley edge states* of the Hermitian model can be regarded as special cases of the PT symmetric edge states analyzed above. We will now compare the representative cases of $m_i = 0$ and $m_r = 0$ in more detail. Figure 4 presents the tight-binding calculations for these two cases in the whole Brillouin zone. Figure 5 schematically shows the results of the ***kp*** approximation in the vicinity of the Dirac points for the case of a locally P-symmetric domain wall (for details, see Appendix C). The calculation demonstrates that both these cases inherit the general characteristics of PT edge states, with the gap either in real [$m_i = 0$, panels (a,c) of Figs. 4,5] or imaginary [$m_r = 0$, panels (b,d)] part of the bulk spectrum crossed by the edge states.

In particular, for $m_i = 0$ Eqs. (12) yield the solutions

$$\epsilon_e = \begin{cases} \pm 1 - \dfrac{m_r - \sqrt{m_r^2 + 4g_k^2}}{2}, & \text{locally P} - \text{symmetric domain wall,} \\[2mm] \pm 1 - \dfrac{m_r + \sqrt{m_r^2 + 4g_k^2}}{2}, & \text{locally P} - \text{broken domain wall.} \end{cases} \quad (14)$$

Four valley edge states located at the locally P-symmetric (red bands) and locally P-broken (blue bands) domain walls are found, among which two bands with parity +1 (symmetric wave function along the interface) cross the band gap and other two with parity -1 (antisymmetric wave function along the interface) lie at the edges of the bulk spectrum, as seen in Fig. 4(a). From Eq. (14) it follows that at $k = \pi$ the valley edge states have energy $\epsilon_e = \pm 1$ for the locally P-symmetric domain wall, which is a general property of the PT edge states. Near the Dirac points, the valley edge states have the well-known linear dispersion [blue line in Fig. 5(a)]

$$\epsilon_e = \frac{-m_r}{2} \pm vk \quad (15)$$

traversing the gap between the Dirac cones of bulk states [shaded areas in Fig. 5(a)]. Here, $k = k_x - \pi \mp \frac{\pi}{3}$ is the detuning of the wave vector from the Dirac point, $v = \frac{\sqrt{3}}{2}$ is the Fermi velocity and we assume that $m_r$ is small. These valley edge states are associated with the valley Hall effect [53], and they can be gapped from bulk states by increasing the magnitude of $m_r$.

In contrast to the valley edge states, which have been widely explored in the literature, the PT edge states appearing solely due to $m_i$ have not been studied thus far. Though the real bulk spectra are not gapped, the imaginary parts of the bulk bands are discontinuous at 0, and the PT edge bands stay within the plane $\text{Im}(\epsilon_e) = 0$ and connect with the bulk bands through parabolic edge bands, as indicated by solid dots in Fig. 4(b,d) as long as $m_i \leq 3$. There is no connection between the PT edge bands and the bulk bands if

$m_i > 3$, as will be shown in the following section from the analysis of EPs in Fig. 6(a). The energies of the PT edge states can be expressed in the compact form

$$\epsilon_e = \pm\sqrt{1-t},  \qquad (16)$$

$$t = \frac{2g_k^2 + m_i^2 \pm \sqrt{(2g_k^2 + m_i^2)^2 - 4g_k^4}}{2}.$$

where the $\pm$ signs in $t$ correspond to the points in the large and small loops, for locally P-symmetric and locally P-broken cases, respectively. The spectra of the edge states when $m_i = 0.3$ are plotted in red and blue color in Fig. 4(b,d). They form two loops and exactly reproduce the numerical tight-binding calculations. Analysis of the PT-symmetric case with $m_r = 0, |m_i| \ll 1$ near the Dirac point is presented in Fig. 5(b,d). Both the tight-binding calculation in Fig. 4(d), and **kp** results in Fig. 5(d), show that the imaginary part of the complex bulk spectrum has a gap of width $m_i$, that is traversed by the parabolic dispersion of edge states

$$\epsilon_e^2 = m_i \pm 2vk. \qquad (17)$$

The spectrum changes dramatically at the exceptional point $k = \mp m_i/(2v)$, where the gap in the real part of the bulk spectrum vanishes and the edge states exhibit the PT phase transition from complex energy (dotted blue curves in Fig. 5) to real energy (solid blue curves). Interestingly, the group velocity corresponding to the dispersion law Eq. (17) diverges at the EPs, although the concept of group velocity should be used with care in the context of non-Hermitian system [54].

In order to further elucidate the difference between the P-symmetric and PT-symmetric interface states we plot in Fig. 5(d,e) the wave functions of interface states in the **kp** model. The wave functions satisfy the general symmetry considerations established in Eqs. (8,9). Namely, for valley states the wave function envelope is real and monotonously decays from the interface, while the corresponding Bloch function has a certain parity, $\beta = 0, \pi$. For PT edge states the envelope function exhibits damped oscillations with distance *y* from the interface $\propto \exp(ipy - \kappa|y|)$, as shown in Fig. 5(e).

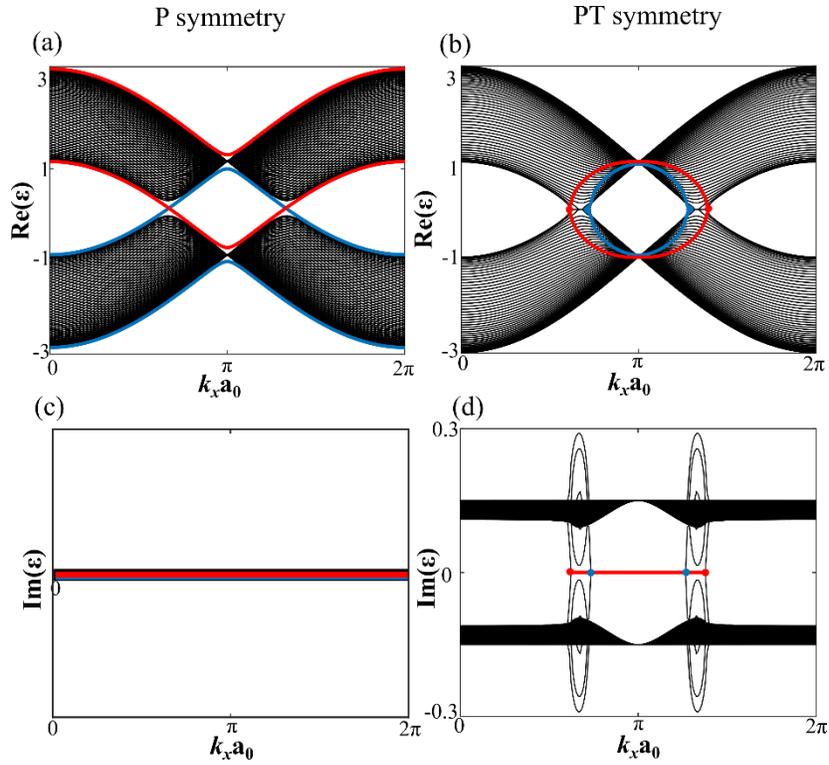

FIG. 4. Comparison of real (a,b) and imaginary (c,d) energy dispersion for two extreme cases, namely, valley edge states with $m_r = 0.3$ and $m_i = 0$, (a,c) and PT edge states with $m_r = 0$ and $m_i = 0.3$ (b,c). Black lines represent the bulk bands, red and blue lines correspond to the edge states at locally P-broken (blue) and P-symmetric (red) interfaces, and the solid dots are the exceptional points. Number of cells employed in the tight-binding method for each domain is $N = 50$.

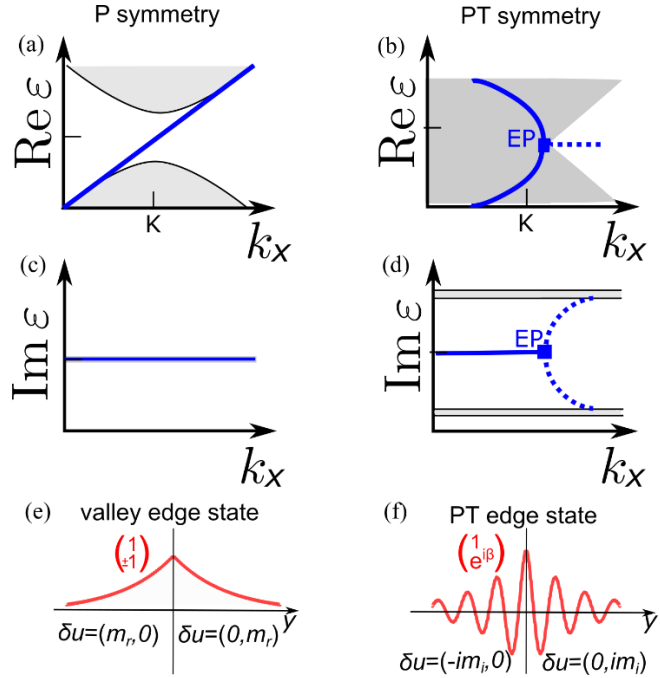

FIG. 5. Same as Fig. 4, but in the ***kp*** approximation near the Dirac point. Panels (a-d) illustrate schematics of two extreme cases, namely, valley edge states with $m_i = 0$ and PT edge states with $m_r = 0$ (b,d). Dispersion of real (a,b) and imaginary (c,d) parts of the complex energies are shown by shaded areas and blue curves for bulk continuum states and PT edge states, respectively. (e,f) Profiles of the real parts of the envelope wavefunction of interface states. Bloch function structure and the on-site potential $\delta u$ for each domain are indicated.

### Effect of local symmetry at the domain wall

Here we examine the effect of the different domain terminations on the existence of edge states. We stress that although Eqs. (13) fully recover the edge states energies, they are obtained without explicit use of the boundary conditions Eqs. (2,3) and rely only on assumption of PT symmetry. Additional insights about the edge states can be drawn from the *local* P-symmetry of the PT-symmetric domain wall, which is preserved in Fig. 1(a) and broken in Fig. 1(b). Since $g(k_x = \pi) = 0$, it follows from Eqs. (1) that the PT edge states at $k_x = \pi$ residing at the domain wall are completely decoupled from the nearest neighbors. This suggests a short decay length $\kappa^{-1} \ll 1$ at $k_x = \pi$, which is verified by the numerical calculation in Fig. S1(a). Consequently, these strongly localized PT edge states only 'see' the local wall symmetry in the red dashed region. The wall in Fig. 1(a) is locally parity-symmetric, and thus the PT edge states can be assigned a certain parity, and their energies derived from Eq. (13) are always equal to $\epsilon_e = \pm 1$ at $k_x = \pi$ no matter what the ratio $r = m_r/m_i$ is. In Fig. 1(b), the parity symmetry for the wall is broken (while the global PT-symmetry is still preserved). Consequently, the PT edge states do not necessarily have a certain parity. The existence of an edge state with energy $\epsilon_e$ depends on the magnitude of $r$ and $m_i$, and the PT edge states vanish if $m_i$ is too large. For example, when $r = 0$, $\epsilon_e = \pm\sqrt{1 - sin(\beta)m_i}$, and if $m_i > sin(\beta)^{-1}$, $\epsilon_e$ becomes complex which contradicts the precondition of PT edge states, so the PT edge states disappear.

With the distinct properties of PT edge states for different domain walls explored, we can easily distinguish between the edge states corresponding to the large loop, which are localized at a locally P-symmetric domain wall, and those corresponding to the small loops, which are localized at a locally P-broken wall for specific parameters $(m_i, m_r)$. The decay length $\kappa^{-1}$ of the PT edge states is calculated from the conditions Eqs. (2) combined with the solutions for different configurations, and is extensively discussed in the Supplementary Material I and III.

### Exceptional points

The case $\beta = \pm\frac{\pi}{2}$ is examined in detail here. Since the Hamiltonian is non-Hermitian, $H \neq H^\dagger$, the right eigenstate $|\psi_e^R(k)\rangle$ and the left eigenstate $|\psi_e^L(k)\rangle$ have to be defined separately to satisfy the eigenvalue equations

$$H(k)|\psi_e^R(k)\rangle = \epsilon_e|\psi_e^R(k)\rangle,$$
$$H^\dagger(k)|\psi_e^L(k)\rangle = \epsilon_e|\psi_e^L(k)\rangle, \quad (18)$$

where $\epsilon_e$ is the eigenenergy of the edge states, which is real. The eigenstates $|\psi_e^{R/L}(k)\rangle$ are given explicitly by

$$|\psi_e^{R/L}(k)\rangle = \sum_{s,n,j} \psi_{e,s,j}(n) |u_{s,j,n}^{R/L}(k)\rangle, \quad n = 0,1,\ldots,N \quad (19)$$

Using the normalization condition $\langle u_i^L(k)|u_j^R(k)\rangle = \delta_{ij}$ [55], and the fact that the vectors $(|u_j^R(k)\rangle, |u_j^L(k)\rangle)$ form a complete basis in the Hilbert space (dual space), the norm of the edge eigenstates is

$$\langle\psi_e^L(k)|\psi_e^R(k)\rangle = \sum_n (e^{i2\beta} + 1)(|\psi_{e,I,A}(n)|^2 + |\psi_{e,I,B}(n)|^2), \quad n = 0,1,\ldots,N. \tag{20}$$

Here, we have exploited the PT-symmetry condition Eq. (7) and the phase factor $\beta$ defined in Eq. (8). Therefore, if $e^{i\beta} = \pm i$, then $\langle\psi_e^L(k)|\psi_e^R(k)\rangle = 0$. The vanishing of the norm indicates that the eigenstates are no longer linearly independent, while having the same eigenvalues. This is the condition for an EP, which is distinct from the case of a band degeneracy [7]. Therefore, the two dispersion curves of PT edge states coalesce at EPs when $\beta = \pm\frac{\pi}{2}$, with the PT symmetric phase being spontaneously broken.

Now we examine the dependence of the position of EPs on the gain/loss parameter $m_i$ based on the discussion above. From Eq. (12) one finds that $\epsilon_e = \pm r$ if $\beta = \pm\frac{\pi}{2}$. Since the EPs of the PT edge states cannot be at $k_x = \pi$, we obtain $\beta = \frac{-\pi}{2}$ for real $\beta = \frac{\pi}{2}$ for locally P-broken domain walls, thus

$$g_k = \begin{cases} \pm\sqrt{(r^2+1)(m_i+1)}, & \text{locally P - preserved domain wall;} \\ \pm\sqrt{(r^2+1)(1-m_i)}, & \text{locally P - broken domain wall.} \end{cases} \tag{21}$$

Therefore, using the condition $0 \leq g_k^2 \leq 4$, we find that the EPs stay near the Dirac points if $r = 0, m_i \to 0$. In another words, the PT symmetry of the modes near the Dirac points is most easily broken compared to modes at other $k$ in the Brillouin zone. This is generally true for bulk modes of the zigzag cut structure because the imaginary part of the complex bulk frequency abruptly changes at the Dirac points due to the perturbation of gain/loss. For a locally P-symmetric domain wall, if $m_r$ is fixed and $m_i$ is continuously increased from 0, the PT edge states first form two separate continuous dispersion curves along the $k_x$ direction, then the EPs of edge states appear at $k_x = 0$ or $2\pi$ and move toward $k_x = \frac{2\pi}{3}$ and $\frac{4\pi}{3}$ and two edge state dispersion curves form a loop during this transition. The EPs are the transition points connecting PT edge states and the complex-valued edge states. The latter have parabolic dispersion curves and link the PT edge bands with bulk bands. Before reaching the Dirac point, the EPs recede back to $k_x = 0$ and $2\pi$, and completely vanish after $m_i$ is tuned to make $(r^2+1)(m_i+1) > 4$. This phase transition in the position of the EPs is shown in Fig. 6(a). In the light red shaded region, no EPs exist, but the PT edge states at $k = \pi$ do. Therefore, these edge states are continuous along $k_x$ and gapped at $k_x = 0, 2\pi$, and the bulk and edge bands are no longer connected through parabolic edge bands (Fig.2 (a,d)). For large enough values $m_r$, EPs are absent at any $m_i$. In case of a locally P-broken wall, as shown in Fig. 1(b), if $m_r$ is fixed and $m_i$ is continuously increased from 0, the PT edge states first have continuous dispersion curves along $k_x$, then the EPs appear at $k_x = 0$ and $2\pi$ and move toward $k_x = \pi$. If $m_i = 1$, the EPs merge at $k_x = \pi$, which indicates that the edge states have broken PT symmetry, as shown in Fig. 6(b). In fact, complex dispersion of edge states is linear and 'degenerate' at $k_x = \pi$. If $m_i > 1$, the PT edge states completely vanish, corresponding to the light blue region in Fig. 6(b). This is consistent with the absence of edge states shown at the locally P-broken domain wall in Fig. 2 (c-d).

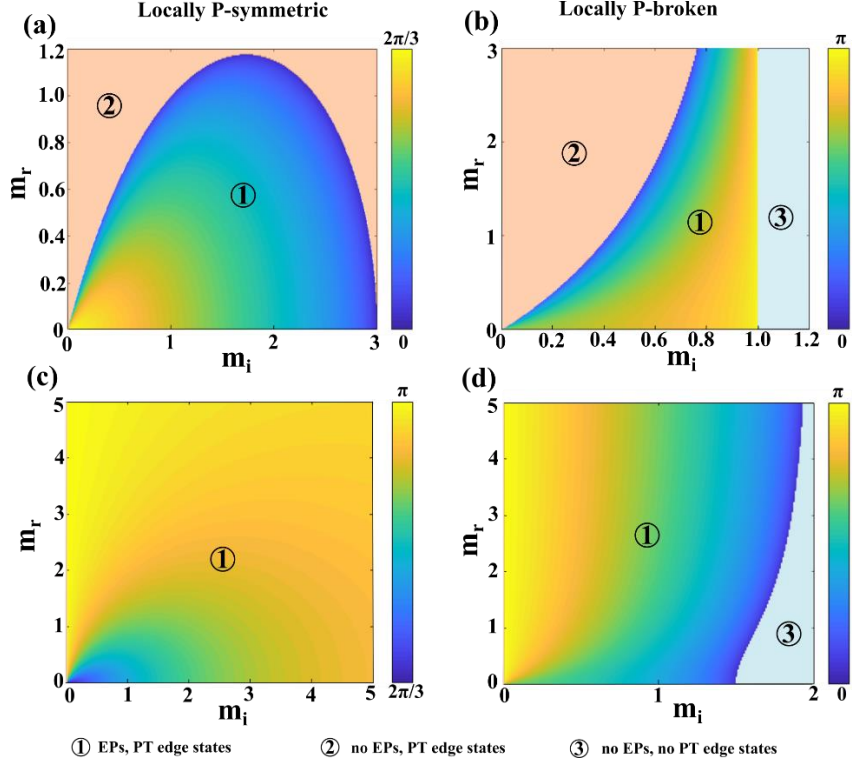

FIG. 6. Phase transition of the EP positions in the Brillouin zone obtained from Eqs. (21,22) depending on $m_i$ and $m_r$ for (a) zigzag cut and locally P-symmetric domain wall, (b) zigzag cut and locally P-broken domain wall, (c) bearded cut and locally P-symmetric and (d) bearded cut and locally P-broken domain wall. Light red and blue shaded colors indicate the regions where the EPs are absent, PT edge states are present in the light red region, but not in the light blue region.

We have also investigated the PT symmetric interfaces with other cuts at the end of the strip, like bearded and armchair cuts, as shown in Fig.1(c-d). The PT edge states are expected to exist in these configurations as well due to the PT symmetry of Hamiltonian. The study of these cases is summarized in the Supplementary Material III. For the bearded cut, the positions of the EPs for locally P-symmetric and locally P-broken domain walls are given by

$$g_k = \begin{cases} \pm\frac{1}{2}\left(m_i - \sqrt{m_i^2 + \frac{4}{r^2+1}}\right), & \text{locally P} - \text{symmetric domain wall,} \\ \pm\frac{1}{2}\left(m_i + \sqrt{m_i^2 + \frac{4}{r^2+1}}\right), & \text{locally P} - \text{broken domain wall.} \end{cases} \quad (22)$$

For a locally P-symmetric domain wall, the EPs move in $k_x$ between $\pm\left(\frac{2\pi}{3}, \pi\right)$ (Fig. 6(c)) and never vanish since $0 < |g_k| < 1$ as long as $m_i \neq 0$. Thus, the loop-shaped dispersion curves of PT edge states located at the locally P-symmetric domain wall always exist, and are robust against the perturbation of the onsite potential and the magnitude of gain/loss. For a locally P-broken domain wall, the EPs exist between wave number $\pm(0, \pi)$ and disappear if $m_i$ is large enough to make $|g_k| > 1$ (Fig. 6(d)). Note that the decay length $\kappa^{-1}$ of the PT edge states for the bearded cut keeps approaching zero within a larger range in $k_x$ if $m_i$ increases, and it is smaller on average than that for the zigzag cut. PT symmetric interfaces for armchair cuts, however, always have two PT edge loops or four gapped PT edge bands that are doubly degenerate in energy, since there is no parity difference between the inner and the outer domain walls. The above analysis indicates that the robustness of PT edge states against the magnitude of gain/loss is for a special feature of the honeycomb lattice, although PT edge states might exist in other lattice structures with PT symmetric interfaces.

Our analysis also clearly demonstrates the importance of local P-symmetry of the domain wall for the system with gain/loss, This symmetry enforces the presence of edge states, and prevents the breaking of global PT-symmetry. The comparison between the valley edge states and PT-edge states for both domain wall configurations is summarized in Table I.

TABLE I. Comparison between P-symmetric and PT-symmetric edge states for two types of domain walls.

|  | P-symmetric | P-broken, PT-symmetric |
|---|---|---|
| Locally P-broken domain wall | Valley edge states | PT edge states for *small* gain/loss |
| Locally P-symmetric domain wall |  | PT edge states for *arbitrary* gain/loss |

Other configurations of gain and loss crystals without PT symmetric interface, schematized in Fig. S4, do not support lossless edge states. Hamiltonians constructed from these configurations are not PT-invariant. Details for different non-PT symmetric interfaces are explained in Supplementary Material IV.

## III. NON-HERMITIAN HALDANE MODEL

The second type of NH model we consider is a Haldane honeycomb lattice consisting of two domains with zigzag cuts at the ends of the strip [56]. Next nearest neighbor (NNN) complex hopping is considered with amplitude $t'$ and phase factor $e^{-i\Phi}$ corresponding to the Haldane flux. In order to construct the PT symmetric interface, we introduce gain at the A sites in domain I, and loss at the B sites in domain II. Periodic boundary conditions are applied along the direction $x_1$ and open boundary conditions are applied at the ends of strip along $x_2$ (Fig. 5). If the magnetic fluxes in domain I and II have the same distribution (Fig. 5(a)), then PT symmetry along the domain wall is destroyed by the local magnetic flux. PT symmetry of the interface can be restored by switching the direction of magnetic fluxes in either one of the domains, as seen in Fig. 5(b). The equations of motion for the two configurations are

$$\epsilon\psi_{I,A}(n) = -h_+\psi_{I,A}(n) - g_-\left(\psi_{I,A}(n+1) + \psi_{I,A}(n-1)\right) - \psi_{I,B}(n+1) - g_0\psi_{I,B}(n) - im_i\psi_{I,A}(n),$$
$$\epsilon\psi_{I,B}(n) = -h_-\psi_{I,B}(n) - g_+\left(\psi_{I,B}(n+1) + \psi_{I,B}(n-1)\right) - \psi_{I,A}(n-1) - g_0\psi_{I,A}(n);$$

$$\epsilon\psi_{II,A}(n) = -h_\pm\psi_{II,A}(n) - g_\mp\left(\psi_{II,A}(n+1) + \psi_{II,A}(n-1)\right) - \psi_{II,B}(n-1) - g_0\psi_{II,B}(n),$$
$$\epsilon\psi_{II,B}(n) = -h_\mp\psi_{II,B}(n) - g_\pm\left(\psi_{II,B}(n+1) + \psi_{II,B}(n-1)\right) - \psi_{II,A}(n+1) - g_0\psi_{II,A}(n) + im_i\psi_{I,A}(n), n = 1,2,\ldots,N-1.$$

(23)

where $h_\pm = 2t'\cos(k \pm \Phi)$, $g_\pm = 2t'\cos(k/2 \pm \Phi)$, $g_0 = 2\cos\left(\frac{k}{2}\right)$. The magnetic fluxes are not present at the domain wall, thus the boundary conditions are

$$\epsilon\psi_{I,A}(0) = -h_+\psi_{I,A}(0) - g_-\left(\psi_{I,A}(1) + \psi_{II,A}(0)\right) - \psi_{I,B}(1) - g_0\psi_{I,B}(0) + m\psi_{I,A}(0),$$
$$\epsilon\psi_{I,B}(0) = -h_-\psi_{I,B}(0) - g_+\left(\psi_{I,B}(1) + \psi_{II,B}(0)\right) - \psi_{II,A}(0) - g_0\psi_{I,A}(0),$$
$$\epsilon\psi_{II,A}(0) = -h_\pm\psi_{II,A}(0) - g_\mp\left(\psi_{II,A}(1) + \psi_{I,A}(0)\right) - \psi_{I,B}(0) - g_0\psi_{II,B}(0),$$

$$\epsilon\psi_{II,B}(0) = -h_{\mp}\psi_{II,B}(0) - g_{\pm}\left(\psi_{II,B}(1) + \psi_{I,B}(0)\right) - \psi_{II,A}(1) - g_0\psi_{II,A}(0) + m^*\psi_{I,A}(0), \quad (24)$$

where $m = im_i$. From the previous analysis, we predict that the PT edge states localized at the domain wall cannot exist in the first configuration shown in Fig. 7(a), but might be present in the second configuration shown in Fig. 7(b) as long as the PT symmetry of the edge states is preserved. The bulk topological invariant of the Haldane model is not changed by introducing the gain/loss into the system, though Berry connection is redefined in the context of NH system. The completeness and orthogonality conditions are only satisfied in the biorthogonal basis [43], and correspondingly Chern number is defined as

$$c = c^{\zeta,\eta} = c^{\eta,\zeta}, \zeta, \eta = R, L, \zeta \neq \eta, \quad (25)$$

where the subscript denotes the right/left basis. It can be shown that the Chern number in Eq. (25) is uniquely defined and is quantized the same way as in the Hermitian context. The details of the gauge transformation and derivation of Berry connection for the NH Haldane model are given in the Appendix B. Based on this, we predict that topological edge states will be present and localized at the ends of the strip and at the domain wall even though their energies might be complex valued.

These predictions are verified by the TBM, and the complex band structures for two configurations are shown in Fig. 8. In both cases all the edge bands are connected with the bulk bands. One-way propagation is also observed for the edge states, revealing the nonreciprocal (chiral) nature of topological edge states in the Haldane model. In the second configuration, PT edge states located at domain wall bridge the gapped bulk bands in the both directions of real and imaginary energies through the two parabolic edge bands (Fig. 8(b)), while PT edge states discussed in the last section bridge gapped bulk bands only in the direction of imaginary energy. If the magnitude of gain/loss is very large, the bulk bands above and below PT edge states in the imaginary energy direction merge with each other, causing the disappearance of topological edge states localized at the ends of chain, while PT edge states at the interface of two domains always survive due to the robustness of PT symmetry phase.

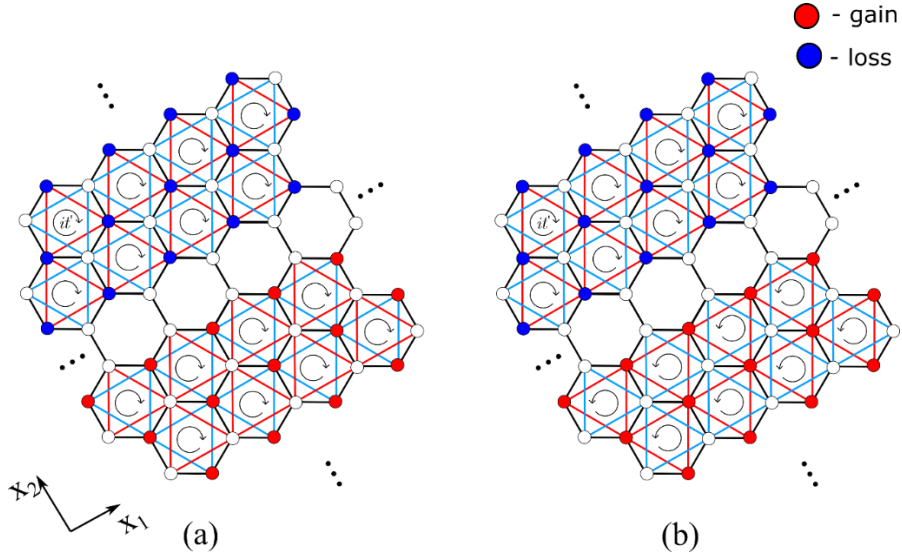

FIG. 7. Nom-Hermitian Haldane model without (a) and with (b) the PT-symmetric interface.

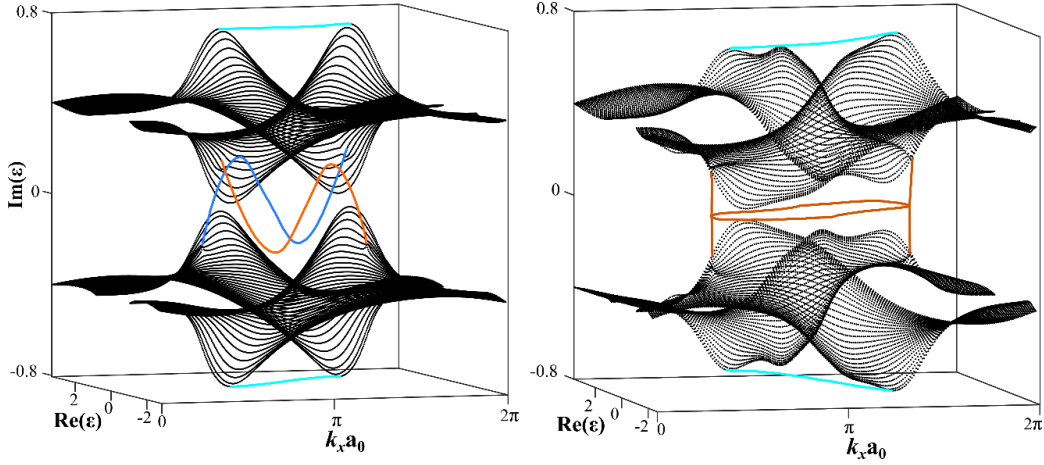

FIG. 8. Complex band structures with $m_i = 0.8, t' = 0.2$, (a) $\phi = \frac{\pi}{3}$ for both domains (b) $\phi = \frac{\pi}{3}$ in the domain I and $\phi = \frac{-\pi}{3}$ in domain II. The cyan curves show the topological edge states located at the ends of the strip, and the red and blue bands in (a) are topological edge states at the domain wall, while the red loop bands in (b) are the topological PT edge states following from topological bulk invariance. Number of unit cells for each domain is $N = 60$.

## IV. OPTICAL IMPLEMENTATION OF PT SYMMETRIC INTERFACES IN PHOTONIC GRAPHENE

PT symmetric systems can be realized in various settings including optical lattices, coupled waveguides, micro resonators and metamaterials [18,19,23,57-60]. To confirm our analytical prediction of PT edge states, we now consider an electromagnetic model relevant to photonics. Specifically, we emulate PT symmetric interfaces in 2D honeycomb photonic crystals composed of dielectric rods (photonic graphene) with the imaginary corrections introduced to the dielectric permittivities of the rods, $\text{Im}(\epsilon_A) = \Delta$ at sites A in domain I and $\text{Im}(\epsilon_B) = -\Delta$ at sites B in domain II.

The effective photonic Hamiltonian near the Dirac points for the photonic crystal with the gain/loss introduced at one site of the unit cell is derived by using the plane-wave expansion of Maxwell's equations (Supplementary Material V)

$$\hat{H}_{K(K')} = \Omega_0 + \delta\Omega_0 \pm V\delta k_x \hat{\sigma}_x + V\delta k_y \hat{\sigma}_y + m\hat{\sigma}_z \quad (26)$$

where $\Omega_0 = K^2(\tilde{\epsilon}_0 + \tilde{\epsilon}_1)$ stands for the unperturbed onsite frequency, $\delta\Omega_0$ denotes the complex correction of the onsite energy, $m$ is the complex mass term due to gain/loss of the material, and $V = K(\tilde{\epsilon}_0 + \tilde{\epsilon}_1)$ is the Fermi velocity. We list the values of $m$ and $\delta\Omega_0$ for different configurations of the unit cell in Table I. Among them, the crystals $\epsilon_A = \epsilon_1 \mp i\Delta, \epsilon_B = \epsilon_1$ and $\epsilon_A = \epsilon_1, \epsilon_B = \epsilon_1 \pm i\Delta$ are PT-symmetric partners.

As follows from Eq. (24), the band degeneracy at the Dirac point is slightly lifted due to the real part of the mass term $m_r$ being of order $\Delta^2$ and inducing inversion symmetry breaking in the unit cell. Moreover, the bulk bands become flattened near the Dirac point due to the imaginary part $m_i \propto \Delta$. These peculiar properties, not observed in Hermitian systems, are confirmed by both tight-binding and plane-wave expansion calculations (Fig. S7). Therefore, photonic lattices with a PT symmetric interface exhibit an effective onsite perturbed potential $\propto \Delta^2$ at sites A in domain I and at sites B in domain II. This corresponds to the model discussed in Section II with $m_r < m_i$ (see Fig. 2 (c)). To model the non-PT symmetric interface, we build the structure in such a way that $\text{Im}(\epsilon_A) = \Delta$ in domain I and $\text{Im}(\epsilon_A) = -\Delta$ in domain II, which corresponds to the configuration discussed in Supplementary Material IV.

Table II. Complex frequency correction $\delta\Omega_0$ and mass term $m$ in the effective Hamiltonian due to gain, loss, and inversion symmetry breaking. $M, M' \propto \Delta$.

|  | $\epsilon_A = \epsilon_1 - i\Delta, \epsilon_B = \epsilon_1$ | $\epsilon_A = \epsilon_1 + i\Delta, \epsilon_B = \epsilon_1$ | $\epsilon_A = \epsilon_1, \epsilon_B = \epsilon_1 - i\Delta$ | $\epsilon_A = \epsilon_1, \epsilon_B = \epsilon_1 + i\Delta$ |
|---|---|---|---|---|
| $m$ | $\left(\dfrac{\Delta}{\epsilon_1} - i\right)M$ | $\left(\dfrac{\Delta}{\epsilon_1} + i\right)M$ | $\left(\dfrac{-\Delta}{\epsilon_1} + i\right)M$ | $-\left(i + \dfrac{\Delta}{\epsilon_1}\right)M$ |
| $\delta\Omega_0$ | $\left(\dfrac{\Delta}{\epsilon_1} + i\right)M'$ | $\left(\dfrac{\Delta}{\epsilon_1} - i\right)M'$ | $\left(\dfrac{\Delta}{\epsilon_1} + i\right)M'$ | $\left(\dfrac{\Delta}{\epsilon_1} - i\right)M'$ |

While the effective $kp$ Hamiltonian Eq. (26) accurately describes the bulk dispersion in the vicinity of the Dirac points, it requires corrections that are quadratic in $\delta k$ to reproduce the dispersion of the PT edge states. This is in stark contrast to the valley edge states, which are captured already by a linear-in-$\delta k$ Hamiltonian. In Appendix (C) we present a rigorous derivation of the effective $kp$ Hamiltonian with $\delta k^2$ terms from the tight-binding method and establish the correspondence between $kp$ and tight-binding considerations of PT and valley edge states near the Dirac point, discussed in Sec. II. The $\delta k^2$ corrections to the Hamiltonian Eq. (26) can be straightforwardly derived from the plane-wave expansion in the same fashion.

The full-wave simulations of electromagnetic response of the photonic crystal supercells with different cuts at the interfaces (zigzag, bearded and armchair) are performed using a finite-element method (FEM) solver (COMSOL Multiphysics). Periodic boundary conditions are imposed in $x_1$ and $x_2$ directions of the supercell, with domains I and II in the lower and upper regions, respectively (Fig. 9, left panel). Thereby, two PT symmetric interfaces are simultaneously present in the geometry. Results of first-principle simulations are summarized in Fig. 9.

First, we model PT and non-PT interfaces with zigzag cuts at the boundaries. In the middle panel of Fig. 9(a) the lossless loop bands (blue color) centered at $k_x = \pi/a_0$ are observed, and these PT edge states are localized at the locally P-symmetric domain wall only, as shown in the left panel. The magnitude of gain/loss $\Delta$ is chosen large enough to make EPs of the PT edge states located at the locally P-broken domain wall disappear, but not large enough to separate the loop bands for PT edge states located at locally P-symmetric domain wall.

Second, for the bearded locally P-symmetric and locally P-broken interfaces, PT edge states with large and small loop bands are observed centered at $k_x = 0$, as seen in Fig. 9(b). We notice that the edge modes at the bearded cut generally decay faster away from the domain wall than those at the zigzag cut. This property is mentioned in Section II and discussed in detail in Supplementary Material I and Supplementary Material III.

Third, two lossless loop bands are found at the armchair PT interfaces and localized at both the domain walls, since the domain walls in this geometry locally have no parity difference.

For all three non-PT symmetric interfaces, no PT edge states in the bandgaps of bulk modes are found, as seen in the right panel of Fig. 9. Thus, our numerical results are consistent with the tight-binding calculations and analytical predictions.

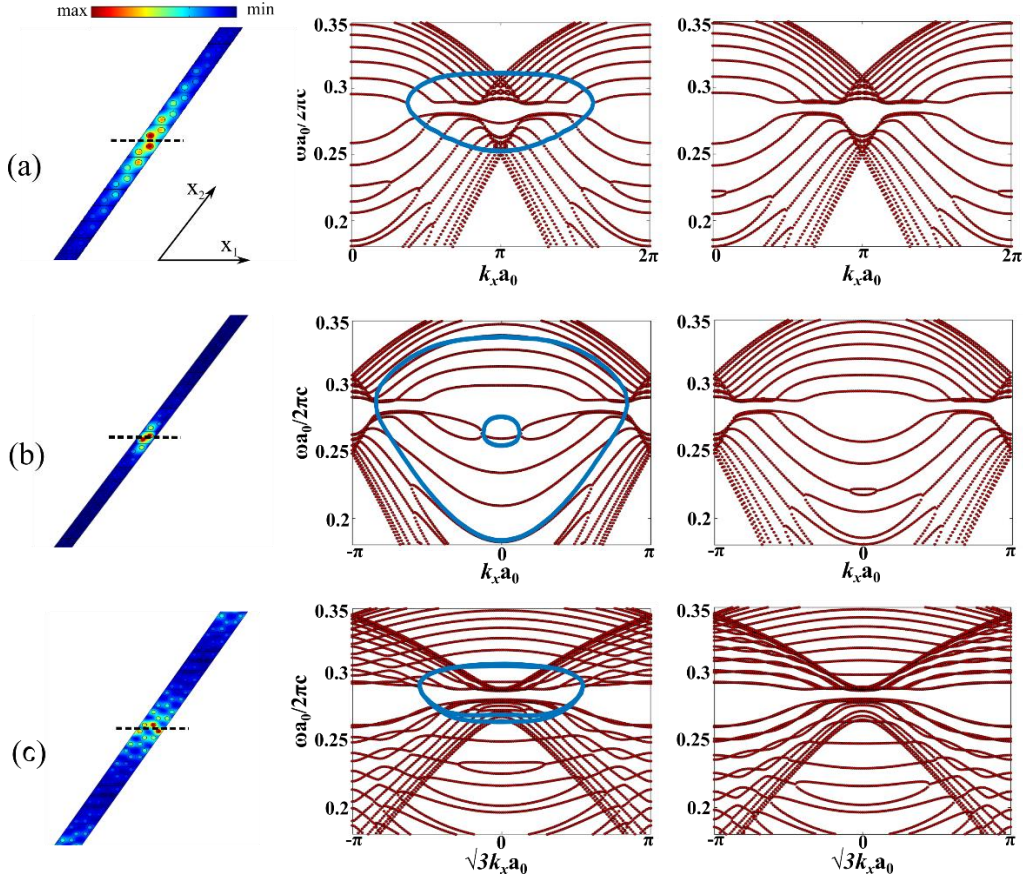

FIG. 9. Optical implementation of PT interfaces in photonic graphene with different cuts at the interfaces (indicated by black dash line): (a) zigzag, (b) bearded, and (c) armchair shaped boundaries. Left panels: Normal electric field $|E|$ profiles for the edge modes localized at different cuts of interfaces between photonic crystals with gain and loss. Middle panels: dispersion (the real part of frequency) for PT-symmetric domain walls. Right panels: dispersion for non-PT domain walls. Branches of PT edge states and bands of dissipative bulk modes are shown in blue and red, respectively. The crystals are made of dielectric rods of radius $r_a = r_b = 0.15a_0$ with permittivity $\epsilon_1 = 14$ and gain/loss parameter $\Delta = 5$ embedded in air. $\epsilon_1 = 14, \Delta = 5$.

## V. CONCLUSIONS

In this paper, we have demonstrated that PT symmetric interfaces between 'gain' and 'loss' honeycomb lattice domains support 'PT edge states' that have real energies. These edge states exist for different cut orientations and domain terminations. We presented a rigorous symmetry analysis unifying the conventional (Hermitian) valley edge states and the edge states of the PT-symmetric structure. Importantly, this analysis revealed that the presence of PT edge states is more sensitive to the local symmetry of the domain wall than the global symmetry of the structure. If the local parity symmetry at the domain wall is broken, the PT edge states exist only for certain values of the lattice parameters. The existence of these edge states is linked to EPs in the edge band; by tuning the magnitude of the gain/loss, it is possible to annihilate the EPs, so that the edge spectrum becomes no longer real. If the domain wall is locally parity-symmetric, the PT edge states are always present no matter how the system is perturbed by onsite potential or gain/loss. When the EPs annihilate, the edge bands detach from the bulk bands and remain real. In particular, EPs of PT edge states confined at a parity-preserved domain wall with bearded cut are robust again the change of gain/loss and onsite perturbed potential (as long as gain/loss not zero), and thus their bands always form loop that connect to the bulk bands through complex parabolic edge bands. In the limit where the gain/loss goes to zero, the edge states reduce to the extensively-studied valley edge states of the Hermitian graphene model.

To further explore the interplay of non-Hermiticity and topology, we also studied the non-Hermitian Haldane model and demonstrated the robustness of its topological features to the introduction of gain/loss. There exist nonreciprocal PT edge states located at the PT symmetric interface, unlike conventional topological edge states, these PT edge states persist for arbitrarily large magnitudes of gain/loss, and exhibit EPs below a critical value of the gain/loss. Last but not least, experimentally feasible optical analogous of honeycomb lattices with and without PT symmetric interface have been studied using first-principles numerical methods, which confirmed the analytical predictions. This work envisions a generalization of Hermitian topological edge states into the NH topological edge states with real spectra.

## ACKNOWLEDGMENTS

We thank Konstantin Bliokh for helpful discussions. The authors gratefully acknowledge financial support from the National Science Foundation grants CMMI-1537294 and EFRI-1641069. ANP acknowledges support by the Australian Research Council and by the Russian "Basis" Foundation.

## APPENDIX A: DERIVATION OF EQUATIONS OF MOTION

We start our analysis by constructing the non-Hermitian Hamiltonian for the PT preserved system from the tight-binding method (TBM) as

$$H = -t_1 \sum_{<i,j>} c_i^\dagger c_j - \sum_{s=I,II} \sum_i m_{s,i} c_{s,i}^\dagger c_{s,i}, \qquad (A1)$$

where $t_1$ is the nearest-neighbor hopping term, which is assumed to be equal for all sites, $m_{s,i}$ is the onsite perturbation satisfying $m_{I,i} = m_{II,i}^*$, which emulates the PT symmetric interface between the domains, and $c_i^\dagger (c_i)$ is the creation (annihilation) operator for boson or fermion at site $i$.

We first consider a graphene nanoribbon consisting of gain (loss) at site $A$ ($B$) in domain I and loss (gain) at site $B(A)$ in domain II, with a length $L = Ma_0$ along the $x$ direction, a finite width $2Na_0$ along the $y$ direction and zigzag cut at the ends, where $a_0$ is the lattice constant. A perturbing potential is also introduced for sites $A(B)$ and size $B(A)$ in domain I and domain II, respectively, as shown in Fig. 1(a) in the main text. We apply Eq. (A1) to this system and define the creation operator,

$$c_{s,j,n}^\dagger(x) = \frac{1}{\sqrt{M}} \sum_k c_{s,j,n}^\dagger(k) e^{ikx}, \qquad (A2)$$

where $k = \frac{2\pi}{L} m, m = 0, \pm 1, \ldots, \pm \frac{M}{2}, s, j$ and $n$ are the index of the particle position, $s = I, II, j = A, B$, and $n = 0,1,2, \ldots, N$. $c_{s,j,n}^\dagger(k)$ is the momentum representation of the creation operator. Thus, Eq. (A1) can be expressed in terms of $c_{s,j,n}^\dagger(k)$. The wave function for the supercell indicated by the black dashed rectangular region in Fig. 1 can be expressed as

$$|\Psi(k)\rangle = \sum_{s,j,n} \psi_{s,j}(n) c_{s,j,n}^\dagger(k) |G\rangle, \qquad (A3)$$

where $|G\rangle$ is the ground state of the Hamiltonian. We solve the eigenvalue problem by plugging Eq. (A3) into the Schrödinger equation

$$H|\Psi(k)\rangle = \epsilon|\Psi(k)\rangle. \tag{A4}$$

As a result, we obtain the equations of motion, as shown in the main text Eqs. (1-3).

**APPENDIX B: TOPOLOGY IN THE NON-HERMITIAN HALDANE MODEL**

For the honeycomb lattice with gain and loss on sites A and B, respectively, the NH Haldane Hamiltonian in momentum space is

$$H(\mathbf{k}) = h_0 + \mathbf{h}\hat{\boldsymbol{\sigma}}$$

$$h_0 = -2t'\cos(\Phi)(\cos(k_x) + \cos\left(-\tfrac{1}{2}k_x + \tfrac{\sqrt{3}}{2}k_y\right) + \cos\left(\tfrac{1}{2}k_x + \tfrac{\sqrt{3}}{2}k_y\right)),$$

$$h_x = -\cos\left(\tfrac{1}{\sqrt{3}}k_y\right) - \cos\left(\tfrac{1}{2}k_x + \tfrac{1}{2\sqrt{3}}k_y\right) - \cos\left(\tfrac{1}{2}k_x - \tfrac{1}{2\sqrt{3}}k_y\right),$$

$$h_y = -\sin\left(\tfrac{1}{\sqrt{3}}k_y\right) + \sin\left(\tfrac{1}{2}k_x + \tfrac{1}{2\sqrt{3}}k_y\right) - \sin\left(\tfrac{1}{2}k_x - \tfrac{1}{2\sqrt{3}}k_y\right),$$

$$h_z = -2t'\sin(\Phi)\left(\sin(k_x) + \sin\left(-\tfrac{1}{2}k_x + \tfrac{\sqrt{3}}{2}k_y\right) + \sin\left(\tfrac{1}{2}k_x + \tfrac{\sqrt{3}}{2}k_y\right)\right) - im_i. \tag{A5}$$

where $\hat{\boldsymbol{\sigma}} = (\sigma_x, \sigma_y, \sigma_z)$ represents the vector of Pauli matrices, $t'$ is the next nearest neighbor (NNN) hopping amplitude, $\Phi$ is the phase of the NNN hopping, which is due to the effect of a local magnetic flux that averages to zero over one unit cell. The dispersive relation of the bulk bands is

$$\delta\epsilon_\pm = \epsilon_\pm - \epsilon_0 = \pm\sqrt{A(k) + h_z^2}, \tag{A6}$$

where $A(k) = h_x^2 + h_y^2 = 3 + 2(\cos(k_x) + \cos\left(\tfrac{1}{2}k_x + \tfrac{\sqrt{3}}{2}k_y\right) + \cos\left(-\tfrac{1}{2}k_x + \tfrac{\sqrt{3}}{2}k_y\right))$, $\epsilon_0 = -2t'\cos(\Phi)A(k)$. Notice that the argument range of $\delta\epsilon_+$ is $\left[\tfrac{-\pi}{2}, \tfrac{\pi}{2}\right)$ and for $\delta\epsilon_-$ is $\left[\tfrac{\pi}{2}, \tfrac{3\pi}{2}\right)$ in order to fix the multivaluedness of the square root function. The eigenvalue equations are written as

$$H(\mathbf{k})|u^R(k)\rangle = \epsilon_\pm|u^R(k)\rangle,$$
$$H(\mathbf{k})^\dagger|u^L(k)\rangle = \epsilon_\pm^*|u^L(k)\rangle. \tag{A7}$$

There are two choices for both right and left eigenstates of the bands, with the completeness condition

$$\langle u^L_{i,n}|u^R_{j,m}(k)\rangle = \delta_{ij}\delta_{nm}, i,j = 1,2, n,m = \pm. \tag{A8}$$

The normalized basis functions can be written in the following form:

$$|u^R_{1,\pm}(k)\rangle = \frac{1}{a(k)}\begin{pmatrix} h_x - ih_y \\ \delta\epsilon_\pm - h_z \end{pmatrix}, |u^R_{2,\pm}(k)\rangle = \frac{1}{b(k)}\begin{pmatrix} \delta\epsilon_\pm + h_z \\ h_x + ih_y \end{pmatrix};$$

$$|u^L_{1,\pm}(k)\rangle = \frac{1}{a(k)^*}\begin{pmatrix} h_x - ih_y \\ \delta\epsilon_\pm^* - h_z^* \end{pmatrix}, |u^L_{2,\pm}(k)\rangle = \frac{1}{b(k)^*}\begin{pmatrix} \delta\epsilon_\pm^* + h_z^* \\ h_x + ih_y \end{pmatrix}, \tag{A9}$$

where $a(k) = \left(2\delta\epsilon_\pm(\delta\epsilon_\pm - h_z)\right)^{\frac{1}{2}}$, and $b(k) = \left(2\delta\epsilon_\pm(\delta\epsilon_\pm + h_z)\right)^{\frac{1}{2}}$ are the normalized factors. At the Dirac points $K_{p/p'} = \frac{4\pi}{3}(\pm 1, 0)$, $A(k) = 0$, while $h_z(K_{pp'}) = \mp\sqrt{3}t'\sin(\Phi) - im_i$. We assume $\sqrt{3}t'\sin(\Phi) > 0$, therefore, at $K_{p/p'}$, $\delta\epsilon_+ = \sqrt{3}t'\sin(\Phi) \pm im_i$, and $\delta\epsilon_- = -\sqrt{3}t'\sin(\Phi) \mp im_i$.

Consequently, the band $\delta\epsilon_+$ has to use wave function $u_{1,+}^\zeta(k)$ ($\zeta = R, L$) to cover $K_p$ and $u_{2,+}^\zeta(k)$ to patch $K_{p'}$, while band $\delta\epsilon_-$ uses $u_{1,-}^\zeta(k)$ at $K_{p'}$ and $u_{2,-}^\zeta(k)$ at $K_p$, otherwise the wave functions are ill defined at the Dirac points. This result indicates that a gauge transformation is needed to connect $u_{1,\pm}^\zeta(k)$ and $u_{2,\pm}^\zeta(k)$ at the arbitrary boundary $\partial BZ$ in the case of gain/loss. From now on we focus on the band $\delta\epsilon_+$ and omit the index $\pm$,

$$|u_1^R\rangle = e^{-i\phi_r}|u_2^R\rangle,$$
$$|u_1^L\rangle = e^{-i\phi_l}|u_2^L\rangle. \quad (A10)$$

Here, $e^{-i\phi_r} = e^{-i\phi_l} = \frac{h_x - ih_y}{\sqrt{h_x^2 + h_y^2}}$, which is not related to the non Hermitian part. From (A14) we find the overlap integral

$$\langle u_j^R | u_j^R \rangle = \frac{h_x^2 + h_y^2 + |\delta\epsilon \pm h_z|^2}{|2\delta\epsilon(\delta\epsilon \pm h_z)|} \geq 1. \quad (A11)$$

Unlike the paper [43], which proposes that the condition $\langle u_j^R | u_j^R \rangle = 1$ always hold, in our model the condition can be satisfied only at the Dirac points. As mentioned in the main text, $(|u_j^R(k)\rangle, |u_j^L(k)\rangle)$ form complete basis sets and have orthogonal condition (A13) in Hilbert basis. We define the Berry connection in the context of NH system using the biorthogonal product:

$$A_1^{\zeta,\eta}(\mathbf{k}) = i\langle u_1^\zeta(\mathbf{k})|\nabla_k u_1^\eta(\mathbf{k})\rangle = e^{\pm i(\phi_\zeta - \phi_\eta)}\left(A_2^{\zeta,\eta}(\mathbf{k}) + \nabla_k \phi_{\eta/\zeta}\right) = \left(A_2^{\zeta,\eta}(\mathbf{k}) + \nabla_k \phi_{\eta/\zeta}\right), \quad (A12)$$

where $\zeta \neq \eta$. The Berry flux is thus calculated as

$$\alpha^{\zeta,\eta} = \oint_{\partial BZ^+} dk A_1^{\zeta,\eta}(\mathbf{k}) + \oint_{\partial BZ^-} dk A_2^{\zeta,\eta}(\mathbf{k}) = \phi_{\eta/\zeta,f} - \phi_{\eta/\zeta,i}. \quad (A13)$$

Since $e^{-i\phi_r} = e^{-i\phi_l}$, it can be concluded that

$$\alpha = \alpha^{\zeta,\eta} = \alpha^{\eta,\zeta}. \quad (A14)$$

Consequently, the topological bulk invariant (Chern number) is uniquely defined and quantized in the NH Haldane model.

### APPENDIX C: k·p MODEL FOR PT EDGE STATES

Although the bulk dispersion in photonic graphene with gain/loss in one sublattice is well captured by the linear-in-$k$ terms of a $\mathbf{k} \cdot \mathbf{p}$ Hamiltonian, it is insufficient to keep only these terms if we wish to derive the dispersion of the PT edge states. Here, we extend the tight-binding analysis to quadratic-in-$k$ terms.

We will focus on the case of purely imaginary mass terms, $m_r = 0$. We write the Schrödinger equations for two domains as

$$\begin{pmatrix} im_i - \varepsilon & h(\delta k_x, \delta k_{y1}) \\ h(\delta k_x, -\delta k_{y1}) & -\varepsilon \end{pmatrix} \begin{pmatrix} \psi_{A1} \\ \psi_{B1} \end{pmatrix} = 0, \quad \text{(A15 a)}$$

$$\begin{pmatrix} -\varepsilon & h(\delta k_x, \delta k_{y2}) \\ h(\delta k_x, -\delta k_{y2}) & -im_i - \varepsilon \end{pmatrix} \begin{pmatrix} \psi_{A2} \\ \psi_{B2} \end{pmatrix} = 0. \quad \text{(A15 b)}$$

Here, the complex-valued column-vectors, are the wave functions in the two domains, corresponding to the complex wavenumbers $\delta k_{y1}$ and $\delta k_{y2}$ in the direction transverse to the domain wall. The function $h(\delta k_x, \delta k_{y1,2})$ is the matrix element of the tight-binding Hamiltonian expanded near the Dirac points up to the quadratic-in-$k$ order.

From Eqs. (A15) and due the PT symmetry, it follows that

$$\begin{pmatrix} \psi_{A1} \\ \psi_{B1} \end{pmatrix} = e^{i\varphi} \begin{pmatrix} \psi_{B2}^* \\ \psi_{A2}^* \end{pmatrix}, \quad \text{(A16)}$$

and $\delta k_{y2} = \delta k_{y1}^*$. The latter condition ensures the continuity of the real part of the wave vector at the interface at $y = 0$, as well as the decay of the wavefunction in the $y$ direction away from the interface. Continuity at the interface also requires that

$$\begin{pmatrix} \psi_{A1} \\ \psi_{B1} \end{pmatrix} = \begin{pmatrix} \psi_{A2} \\ \psi_{B2} \end{pmatrix}. \quad \text{(A17)}$$

Hence, we derive

$$\frac{\psi_{A2}}{\psi_{B2}} = \frac{h(\delta k_x, \delta k_{y1}^*)}{\varepsilon} = \frac{\psi_{B2}^*}{\psi_{A2}^*} = \frac{h(\delta k_x, \delta k_{y1})}{\varepsilon - im_i} = \frac{\psi_{A1}}{\psi_{B1}} = \frac{\varepsilon}{h(\delta k_x, -\delta k_{y1})} = \frac{\psi_{B1}^*}{\psi_{A1}^*} = \frac{\varepsilon + im_i}{h^*(\delta k_x, \delta k_{y1})}. \quad \text{(A18)}$$

where we have substituted $\delta k_{y2} = \delta k_{y1}^*$ and used $h^*(\delta k_x, -\delta k_{y1}^*) = h(\delta k_x, \delta k_{y1})$. The relation between the components of the edge state wavefunction in Eq. (A18) can be interpreted as an effective boundary condition. Remarkably, the case of P-symmetric interface suggests $\psi_{A1} = \pm\psi_{B1}$, and purely imaginary $\delta k_{y1}$. The correspondence between the $\mathbf{k} \cdot \mathbf{p}$ and tight-binding models is recovered through

$$\frac{\psi_{A2}}{\psi_{B2}} = \frac{\psi_{A1}}{\psi_{B1}} = e^{-i\beta} = \frac{a_I}{b_I} = \frac{a_{II}}{b_{II}} = \frac{\psi_{e,II,A}(0)}{\psi_{e,I,B}(0)}. \quad \text{(A19)}$$

From Eq. (A18), we rewrite the boundary condition as the real equation

$$\varepsilon^2 = h(\delta k_x, \delta k_{y1}^*)h(\delta k_x, -\delta k_{y1}), \quad \text{(A20)}$$

and recover the complex dispersion equation

$$\varepsilon(\varepsilon - im_i) = h(\delta k_x, -\delta k_{y1})h(\delta k_x, \delta k_{y1}). \quad \text{(A21)}$$

Here, we denote $\delta k_{y1} = \delta k_y - i\kappa$. Exploiting Eqs. (A20), one may derive the dispersion of the edge states $\varepsilon(\delta k_x)$ near the Dirac point. To get an analytical approximation for $\varepsilon(\delta k_x)$, we consistently assume $\varepsilon^2 \sim \xi^2 \sim \delta k_y^2 \sim \delta k_x \sim \kappa$, where we have formally introduced a small parameter $\xi$. From the tight-binding model near the Dirac point $(4\pi/3, 0)$, keeping only the order $\xi^2$, we find

$$h(\delta k_x, \delta \bar{k}_{y1}) \approx \frac{\sqrt{3}}{2}(\delta k_x - i\delta \bar{k}_{y1}) - \frac{3}{8}\delta \bar{k}_{y1}^2, \quad \text{(A22)}$$

where $\delta \bar{k}_{y1} = \frac{\delta k_{y1}}{\sqrt{3}} \approx \frac{2}{\sqrt{3}}\varepsilon + \delta \bar{k}_y - i\bar{\kappa}$. We then substitute

$$h(\delta k_x, \delta \bar{k}_{y1}) \approx \frac{\sqrt{3}}{2} \left(\delta k_x - i\delta \bar{k}_y - \bar{\kappa}\right) - \frac{\varepsilon^2}{2} - i\varepsilon, \tag{A23}$$

into Eq. (A18) and obtain

$$\frac{3}{4}(\delta k_x + \bar{\kappa})^2 - \frac{\sqrt{3}}{2}\varepsilon^2(\delta k_x + \bar{\kappa}) + \frac{\varepsilon^4}{4} + \sqrt{3}\,\delta\bar{k}_y\varepsilon = 0. \tag{A24}$$

Next, we separate the real and imaginary parts of the dispersion equation (A21). From the equation for the imaginary part we obtain $\bar{\kappa} = \frac{m_i}{\sqrt{3}}$, while the equation for the real part is recast as

$$\frac{3}{4}(\delta k_x^2 - \bar{\kappa}^2) - \frac{\sqrt{3}}{2}\varepsilon^2\delta k_x + \frac{\varepsilon^4}{4} + \sqrt{3}\,\delta\bar{k}_y\varepsilon = 0. \tag{A25}$$

Subtracting Eq. (A25) from Eq. (A24), we get $\varepsilon^2 = m_i + \sqrt{3}\,\delta k_x$, which is in full agreement with Eq. (17).

[1] C. M. Bender and S. Boettcher, Phys. Rev. Lett. **80**, 5243 (1998).
[2] P. Dorey, C. Dunning, and R. Tateo, J. Phys. a-Math. Gen. **34**, 5679 (2001).
[3] A. Mostafazadeh, J. Math. Phys. **43**, 2814 (2002).
[4] C. M. Bender, D. C. Brody, and H. F. Jones, Phys. Rev. Lett. **89** (2002).
[5] M. V. Berry, Czech J. Phys. **54**, 1039 (2004).
[6] W. D. Heiss, J. Phys. a-Math. Gen. **37**, 2455 (2004).
[7] W. D. Heiss, J. Phys. a-Math. Theor. **45** (2012).
[8] B. Zhen, C. W. Hsu, Y. Igarashi, L. Lu, I. Kaminer, A. Pick, S. L. Chua, J. D. Joannopoulos, and M. Soljacic, Nature **525**, 354 (2015).
[9] J. Doppler *et al.*, Nature **537**, 76 (2016).
[10] T. Gao *et al.*, Nature **526**, 554 (2015).
[11] L. Feng, R. E. Ganainy, and L. Ge, Nat. Photonics **11**, 752 (2017).
[12] C. E. Ruter, K. G. Makris, R. El-Ganainy, D. N. Christodoulides, M. Segev, and D. Kip, Nat. Phys. **6**, 192 (2010).
[13] M. C. Zheng, D. N. Christodoulides, R. Fleischmann, and T. Kottos, Phys. Rev. A **82** (2010).
[14] K. G. Makris, R. El-Ganainy, D. N. Christodoulides, and Z. H. Musslimani, Phys. Rev. Lett. **100** (2008).
[15] L. Ge, Y. D. Chong, and A. D. Stone, Phys. Rev. A. **85** (2012).
[16] Z. Lin, H. Ramezani, T. Eichelkraut, T. Kottos, H. Cao, and D. N. Christodoulides, Phys. Rev. Lett. **106** (2011).
[17] A. Regensburger, C. Bersch, M. A. Miri, G. Onishchukov, D. N. Christodoulides, and U. Peschel, Nature **488**, 167 (2012).
[18] L. Feng, Y. L. Xu, W. S. Fegadolli, M. H. Lu, J. E. B. Oliveira, V. R. Almeida, Y. F. Chen, and A. Scherer, Nat. Mater. **12**, 108 (2013).
[19] L. Feng, Z. J. Wong, R. M. Ma, Y. Wang, and X. Zhang, Science **346**, 972 (2014).
[20] H. Hodaei, M. A. Miri, M. Heinrich, D. N. Christodoulides, and M. Khajavikhan, Science **346**, 975 (2014).
[21] A. V. Poshakinskiy, A. N. Poddubny, and A. Fainstein, Phys. Rev. Lett. **117** (2016).
[22] L. Feng, M. Ayache, J. Q. Huang, Y. L. Xu, M. H. Lu, Y. F. Chen, Y. Fainman, and A. Scherer, Science **333**, 729 (2011).
[23] B. Peng *et al.*, Nat. Phys. **10**, 394 (2014).
[24] M. Wimmer, A. Regensburger, M. A. Miri, C. Bersch, D. N. Christodoulides, and U. Peschel, Nat. Commun. **6** (2015).
[25] M. C. Rechtsman, J. M. Zeuner, Y. Plotnik, Y. Lumer, D. Podolsky, F. Dreisow, S. Nolte, M. Segev, and A. Szameit, Nature **496**, 196 (2013).

[26] A. B. Khanikaev, S. H. Mousavi, W. K. Tse, M. Kargarian, A. H. MacDonald, and G. Shvets, Nat. Mater. **12**, 233 (2013).
[27] K. J. Fang, Z. F. Yu, and S. H. Fan, Nat. Photonics **6**, 782 (2012).
[28] M. Hafezi, E. A. Demler, M. D. Lukin, and J. M. Taylor, Nat. Phys. **7**, 907 (2011).
[29] Z. Wang, Y. D. Chong, J. D. Joannopoulos, and M. Soljacic, Nature **461**, 772 (2009).
[30] S. Raghu and F. D. M. Haldane, Phys. Rev. A **78** (2008).
[31] F. D. M. Haldane and S. Raghu, Phys. Rev. Lett. **100** (2008).
[32] B. A. Bernevig and T. L. Hughes, *Topological insulators and topological superconductors* (Princeton University Press, Princeton, 2013).
[33] X. L. Qi and S. C. Zhang, Rev. Mod. Phys. **83** (2011).
[34] M. Z. Hasan and C. L. Kane, Rev. Mod. Phys. **82**, 3045 (2010).
[35] L. Lu, J. D. Joannopoulos, and M. Soljaclc, Nat. Photonics **8**, 821 (2014).
[36] J. M. Zeuner, M. C. Rechtsman, Y. Plotnik, Y. Lumer, S. Nolte, M. S. Rudner, M. Segev, and A. Szameit, Phys. Rev. Lett. **115** (2015).
[37] M. S. Rudner and L. S. Levitov, Phys. Rev. Lett. **102** (2009).
[38] H. Schomerus, Opt. Lett. **38**, 1912 (2013).
[39] T. E. Lee, Phys. Rev. Lett. **116** (2016).
[40] S. Weimann, M. Kremer, Y. Plotnik, Y. Lumer, S. Nolte, K. G. Makris, M. Segev, M. C. Rechtsman, and A. Szameit, Nat. Mater. **16**, 433 (2017).
[41] K. Esaki, M. Sato, K. Hasebe, and M. Kohmoto, Phys. Rev. B **84** (2011).
[42] D. Leykam, K. Y. Bliokh, C. L. Huang, Y. D. Chong, and F. Nori, Phys. Rev. Lett. **118** (2017).
[43] H. T. Shen, B. Zhen, and L. Fu, arXiv:1706.07435 (2017).
[44] A. H. Castro Neto, F. Guinea, N. M. R. Peres, K. S. Novoselov, and A. K. Geim, Rev. Mod. Phys. **81**, 109 (2009).
[45] Y. Plotnik *et al.*, Nat. Mater. **13**, 57 (2014).
[46] T. Ma and G. Shvets, New. J. Phys. **18** (2016).
[47] H. Ramezani, T. Kottos, V. Kovanis, and D. N. Christodoulides, Phys. Rev. A **85** (2012).
[48] A. Szameit, M. C. Rechtsman, O. Bahat-Treidel, and M. Segev, Phys. Rev. A **84** (2011).
[49] Y. C. Hu and T. L. Hughes, Phys. Rev. B **84** (2011).
[50] G. Q. Liang and Y. D. Chong, Phys. Rev. Lett. **110** (2013).
[51] G. Harari, Y. Plotnik, M. A. Bandres, Y. Lumer, M. Rechtsman, and M. Segev, CLEO: QELS_Fundamental Science 2015 (2015).
[52] G. W. Semenoff, V. Semenoff, and F. Zhou, Phys. Rev. Lett. **101** (2008).
[53] D. Xiao, W. Yao, and Q. Niu, Phys. Rev. Lett. **99** (2007).
[54] H. Schomerus and J. Wiersig, Phys. Rev. A **90** (2014).
[55] D. C. Brody, J. Phys. a-Math Theor. **47** (2014).
[56] F. D. M. Haldane, Phys. Rev. Lett. **61**, 2015 (1988).
[57] A. Guo, G. J. Salamo, D. Duchesne, R. Morandotti, M. Volatier-Ravat, V. Aimez, G. A. Siviloglou, and D. N. Christodoulides, Phys. Rev. Lett. **103** (2009).
[58] Y. L. Xu, W. S. Fegadolli, L. Gan, M. H. Lu, X. P. Liu, Z. Y. Li, A. Scherer, and Y. F. Chen, Nat. Commun. **7** (2016).
[59] J. S. Tang *et al.*, Nat. Photonics **10**, 642 (2016).
[60] M. Lawrence, N. N. Xu, X. Q. Zhang, L. Q. Cong, J. G. Han, W. L. Zhang, and S. Zhang, Phys. Rev. Lett. **113** (2014).